\documentclass[a4paper]{jpconf}
\usepackage{graphicx}
\usepackage{amsmath}
\newcommand{\grad}{$^{\circ}$}
\begin{document}
\title{Study of Risetime as a function of the distance to the Shower Core in the Surface Detector (SD) of the Pierre Auger Observatory}

\author{Hern\'an Castellanos Vald\'es $^{a}$, Karen Salom\'e Caballero Mora $^{a,b}$ }
\address{$^{a}$Universidad Aut\'onoma de Chiapas;  $^{b}$Mesoamerican Centre for Theoretical Physics (MCTP)}

\ead{hernancastellanosvaldes@gmail.com}

\begin{abstract}
 Cosmic Rays (\textbf{CR}) are high energy particles which come from the universe. When one of those particles enters to the earth's atmosphere it produces an air shower, conformed by secondary particles in which the initial energy is distributed. The Pierre Auger Observatory, located in Argentina, is dedicated to the study of those events. One of the main goals is to find out where those \textbf{CR} are coming from and which kind of chemical composition do they have. In this work we show the status of a study of the risetime ($t_{1/2}$) as a function of the distance to the shower core (near to the air shower's axis) for different zenith angles and energies, obtaining a new variable that will  be compared with other variables used by the Observatory. 
The main objective of this study is to better understand risetime as a mass composition sensitive parameter of \textbf{CR}.
\end{abstract}

\section{Introduction}
Ultra High Energy Cosmic Rays (UHECR) are high energy particles that come from the universe. Due to their low frequency they can only be studied indirectly when one of these particles enters to the Earth's atmosphere and it produces an air shower, conformed by secondary particles in which the initial energy is distributed \cite{a}.
\subsection{Pierre Auger Observatory}
The Pierre Auger Observatory, located in Argentina, is dedicated to the study of those events. One of the main goals is to find out where those UHECR are coming from and which kind of chemical composition do they have. The Observatory has two kind of detectors, the Surface Detector (SD), and 24 Fluorescence Detectors (FD) \cite{b}. The Observatory is also equipped with detectors for atmospheric monitoring and hosts a radio array for the study of EAS at MHz frequencies among other upgrades recently deployed and in process to be deployed \cite{c}.
\subsection{Surface Detector}
The Surface Detector covers an area of 3000 km$^{2}$ with 1660 water Cherenkov detectors , with 12000 l ultrapurified water and three Photomultiplier Tubes each one.
\newpage
\section{Offline}
The Offline \cite {d} software framework has been developed within the Pierre Auger Collaboration to perform the computing tasks. It is designed to provide all the functionalities to process data from the shower detector, including simulation and reconstruction, and it is implemented in C++. The Offline framework provides an easy-to-use interface, making possible  to modify the codes and to include new tasks, completely hiding the detailed internal mechanisms of how and where the data are taken from. There is also a general purpose from SD, FD as well as hybrid event reconstruction. For this work the \textit{advanced data summary trees} ADST data samples produced with Offline are used, to manipulate them only a valid ROOT \cite{e} installation is necessary. Figure 1 shows an event observed by the FD using the Offline graphic tools.
\begin{figure}[h]
\includegraphics[width=25pc]{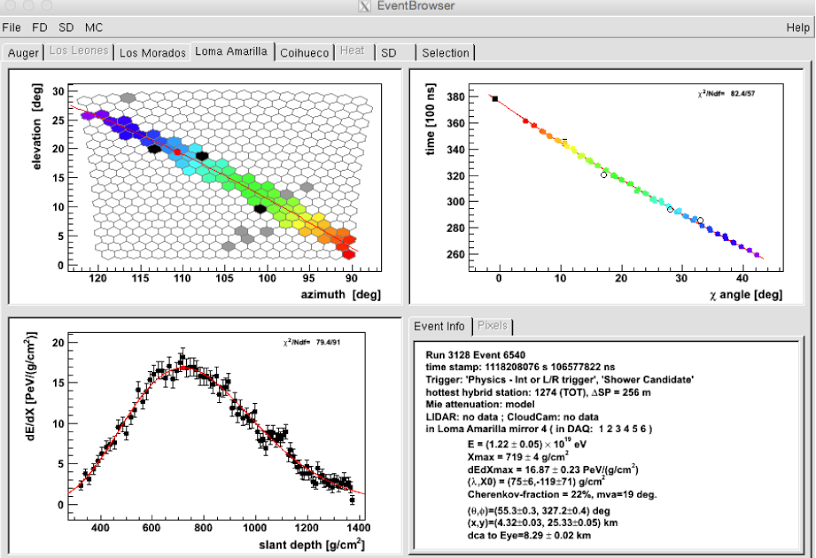}\hspace{.5pc}%
\begin{minipage}[b]{12pc}\caption{\label{offline} An event measured by the FD in the viewer of Offline.}
\end{minipage}
\end{figure}

\section{Risetime }
The $t_{\frac{1}{2}}$  is the time it takes to the integrated signal recorded by the SD to rise from $10 \%$ to $50 \%$ of the final value, a diagram of the risetime for a simulated signal is shown in Figure 2 \cite{f}. It is a mass composition sensitive parameter of UHECR since for different primary particles there is different risetime value. Figure 3  \cite{g} shows that the risetime for iron particles is smaller than that for light particles as proton at a distance of 1000 m from the shower core, from simulated showers. The signal produced by different primaries results in different risetimes due, among other detector effects, to the depth of the first interaction in the atmosphere as shown in Figure 4 \cite{f}. For heavy primaries the interaction occurs early, then the electromagnetic component of the air shower attenuates at ground more than it happens when the interactions occurs deep in the atmosphere as in the case of light particles. Therefore, for the first case the risetime is smaller than for the second one. 
\begin{figure}[ht]
\begin{minipage}{20pc}
\includegraphics[width=20pc]{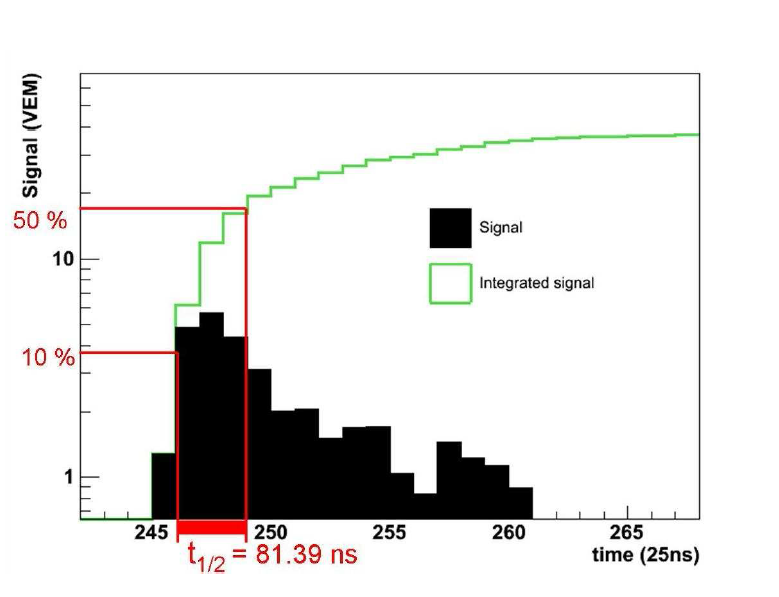}
\caption{\label{rs} Diagram of the definition of $t_{\frac{1}{2}}$  \cite{f}.}
\end{minipage}
\begin{minipage}{19pc}
\includegraphics[width=19pc]{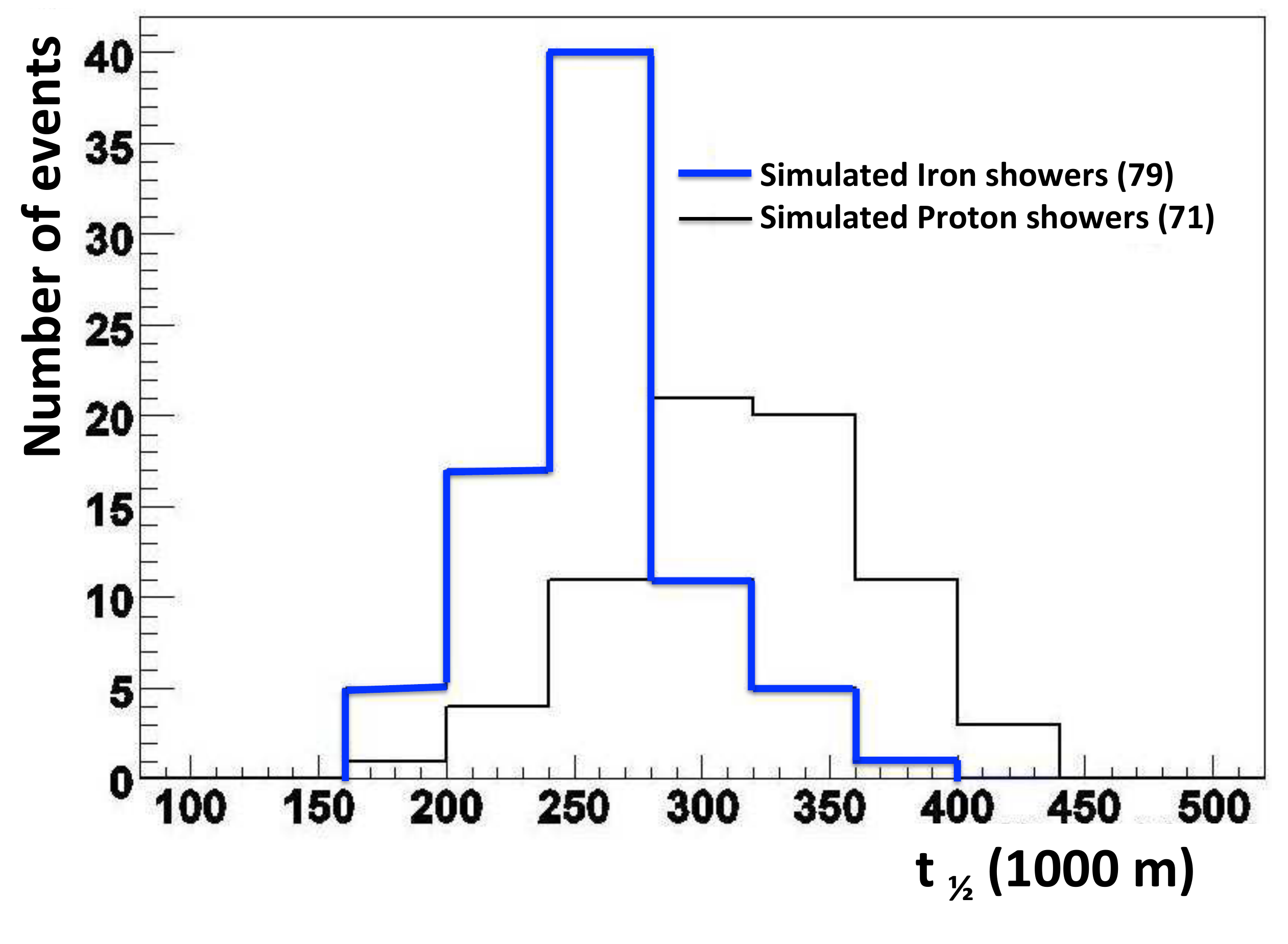}
\begin{minipage}[b]{18pc}\caption{\label{rs1}Comparison of $t_{\frac{1}{2}}$ for heavy and light particles, from simulated showers \cite{g}.}
\end{minipage}
\end{minipage}
\end{figure}
\begin{figure}[ht]
\begin{center}
\begin{minipage}{18pc}
\includegraphics[width=14pc]{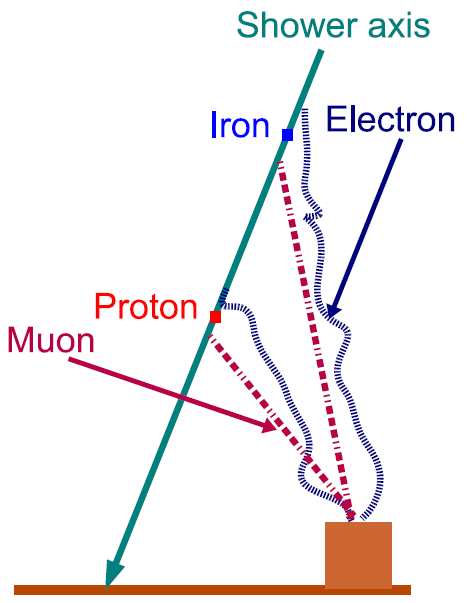}
\begin{minipage}[b]{20pc}\caption{\label{rs3} Diagram of the difference between the interaction of heavy particles and protons with the atmosphere \cite{f}.}
\end{minipage}
\end{minipage}
\end{center}
\end{figure}
\vspace{-1pc}
\section{Work Description}
Typically a value of $t_{\frac{1}{2}}$ is considered for each event, this is the one a detector located at 1000 m from the shower core would have, no matter the energy or zenith angle of the event. Nevertheless this value shows a large spread \cite{f}. This does not allow to directly use this parameter as mass composition sensitive. The main objective of this study is to find a distance to the shower core, zenith angle or energy dependent, for any particular event which might not necessarily be 1000 m. It is expected this parameter presents a smaller spread than the one of $t_{\frac{1}{2}}$ at 1000 m. Once the distance as function of the zenith angle or of the energy is found, the corresponding $t_{\frac{1}{2}}$ will be tested as mass composition sensitive parameter to be compared with other SD parameters. The first attempt is to find the distance as a function of zenith angle for a similar energy.
 
 \noindent Events with angles around 30\grad - 39\grad, 40\grad - 49\grad and 50\grad - 59\grad  are analysed. Fits of the $t_{\frac{1}{2}}$ as a function of shower core for a certain event are performed. The forms of the used fts are the following:
\begin{eqnarray}
\notag f(x)=40+ax+bx^{2} \text{ \cite{f} and \cite{h}} \\
\notag f(x)=10 + \sqrt{a^{2}+bx^{2}}-a \text{ \cite{i}} \\
\notag f(x)=40+ax^{b} 
\end{eqnarray}
In a range of 600 m to 1500 m of distance to the shower core. The intersection of the fits for the same event will be the distance we want to find. This first study shows that there is a different intersection distance of the fitted functions of $t_{\frac{1}{2}}$, for events with different zenith angles. That distance might be used for performing more accurate studies on mass composition. A total of 15 events have been analysed showing similar results. Table 1  shows the preliminary results of three of the analysed events. This common point is found by eye. Currently a method to find the intersection point through a code for several events and setting better quality cuts is being performed.  The dependence of the distance with the energy will be also explored and after finding a good characterization of such intersection distance, studies on mass composition will be performed.
\begin{center}
\begin{table}
  \begin{minipage}[b]{0.98\linewidth}
 \centering
\begin{tabular}{llll}
\br
Zenith Angle &  Energy &InD\\
($\circ$)&x $10^{19}$ (eV)&(m)\\
\mr
36.56&2.29&765\\
44.77&1.37&910\\
57.57&1.62&1095\\
\br
\end{tabular}
\centering
\caption{Examples of characteristic distances for $t_{1/2}$ for different zenith angles and similar energies.}
\end{minipage}
\end{table}
\end{center}
\section{Conclusion}
The $t_{\frac{1}{2}}$, as a mass composition sensitive parameter, could be optimized if the characteristic length for each event is found. the results about composition could improve with respect to the data taken at 1000 m.
\section{Aknowledgements}
\begin{itemize}
\item To CONACyT for the support granted through the project CB 243290
\item To L'OR\'EAL M\'exico, CONACyT, UNESCO, CONALMEX and AMC for the support granted through the "Beca para Mujeres en la Ciencia 2014" for Karen Salom\'e Caballero Mora.
\end{itemize}

\end{document}